# Miniaturization and control of split ring structures from an analytic solution of their resonance

S. T. Chui[1], Y. Zhang[2] and L. Zhou[2]

[1] *Bartol Research Institute and Department of Physics and Astronomy, University of Delaware, Newark, DE 19716,*

[2] *Surface Physics Laboratory (State Key Laboratory) and Physics Department, Fudan University, Shanghai 200433, China*

Abstract

We derived simple polynomial equations to determine the entire resonance spectra of split ring structures. For double stacking split rings made with flat wires, we showed that the resonance frequency depends linearly on the ring-ring separation. In particular, we found that the wavelength of the lowest resonance mode can be made as large as $10^5$ the geometrical size of the ring for realistic experimental conditions, whereas for current systems this ratio is of the order of 10. Finite-difference-time-domain simulations on realistic structures verified the analytic predictions.

## I. Introduction

Stimulated by interests in negative reflecting materials [1-2], there has been much interest in split ring resonators (SRR) [3-7]. A theme in current research is the development of resonators with geometrical sizes much smaller than the wavelengths at resonance. Miniaturization of resonators will open the door to different technological applications such as the subwavelength ultra-compact dipole antennas [8] and filters [9] that have been proposed. Recently, we showed that for two rings of radii $R$ placed on the same plane made by wires possessing a circular cross-section with radius $a$, the resonance frequency exhibits a lower bound $\propto \left[\ln(R/a)\right]^{-0.5}$ [7,10]. To produce a significant change to the slowly changing log function requires a big *reduction of a*, which increases the resistance of the wire and hence the damping of the structure.

In this paper we show how to reduce the resonance frequency of structures made with rings. Our insight is obtained through our recent efforts to understand the resonance properties of metallic ring systems more *rigorously* and *analytically* [7]. In contrast, most previous calculations on metallic rings are numerical in nature. In this paper, we show further that the entire spectra of the SRR systems (both single- and double-ring) can be determined by simple polynomial equations (Eqs. 4 and 11). When the rings are made with flat wires of a film-like rectangular cross-section and placed on different planes [5], their separation $d$ needs only be larger than the thickness $t$ of the rings which can be much smaller than width $2a$ of the wires that make up the ring. The resonance frequency is controlled by the difference between the inverses of the self and the mutual capacitances, which can be very small. For example, it is experimentally possible to make bilayer systems separated by a distance of less than 10 nm. For rings of radii 10 mm, the resonance wavelength can be made $10^5$ the radius of the ring. This huge ratio is previously inaccessible (For current systems, this ratio is 10).

Such a double-stacking SRR was first proposed in Ref. [5] and studied by an empirical method [5], but the attention mainly focused on reducing the bianisotropy of the SRR systems. Here, we employed a more rigorous theory [7] to study the same system, our attention mainly focused on how to lower the resonance frequency with such a system. The extreme limit when the ring radii are much larger than their separation involves the cancellation of two quantities close in values. This requires careful control of the numerical accuracies of the calculation complemented by analytic considerations, which is feasible in the present approach. We now describe our results in detail.

**II. Single Ring**

We first recapitulate our results for the single ring case [7]. Take the coordinate system so that the $z$ axis is perpendicular to the ring and assume a time dependence $e^{i\omega t}$. Driven by an external field, the current along the ring is determined by $\sum_{m'} \rho(m-m') I_{m'} + i I_m \left[ L_m \omega - m^2/(\omega C_m) \right] = E_{ext}(m)$, where $\rho(m), I_m, E_{ext}(m)$ are the Fourier components of the resistivity function $\rho(\phi)$ of the ring, the current $I(\phi)$ flowing on the ring, and the external field $E_{ext}(\phi)$ on the ring [7]. Here, $L_m, C_m$ are the self inductance and capacitance [7], and $\phi$ is the azimuthal angle. The resonance frequencies are obtained by solving the above circuit equation setting $E_{ext}(\phi) = 0$ [7].

For a single-ring SRR with a gap at $\phi = 0$, the resistivity function is $\rho(m) \approx r + \delta_{m,0} r_c$ in the limit of gap width approaching zero; here $r$ and $r_c$ are the resistances per unit length of the insulating gap and the conductor [7]. We take the limit of $r \to \infty$ in the end. When the gap is located at $\phi = \pi$, we get $\rho(m) \approx (-1)^m r + \delta_{m,0} r_c$. In matrix form, the circuit equation to determine the resonance frequencies is

$$\mathbf{H}\mathbf{I} = \mathbf{0}, \tag{1}$$

with

$$\mathbf{H} = \mathbf{H_0} + \mathbf{X}, \tag{2}$$

where $\mathbf{H_0} = r\mathbf{M}$, $\mathbf{M}$ is a matrix with all elements equal to 1, ie. $\mathbf{M}_{i,j} = 1$. $\mathbf{X}$ is a diagonal matrix with $X_m = r_c + i[L_m \omega - m^2/(C_m \omega)]$. For our problem, $X_{-m} = X_m$.

Because of the $m$, $-m$ symmetry of $\mathbf{X}$, there are two classes of solutions, corresponding to even and odd symmetries under the transformation from $m$ to $-m$. Those with odd symmetries (i.e., $I_{-m} = -I_m$) are not coupled to each other and the resonance frequencies are given by the conditions that $X_m = 0$. In the limit of $a/R \to 0$, the resonance frequencies (obtained by solving $X_m = 0$ with $r_c = 0$) are

$$\omega_{2m} = m/\sqrt{L_m C_m} \to m\omega_u, \qquad m = 1, 2, \ldots \qquad (3)$$

with $\omega_u = c/R$ ($c$ is the speed of light) being the frequency unit of the present problem [7]. Equation (3) accurately accounts for all the even-numbered resonance modes of a single-ring SRR obtained numerically (see Fig. 2 of [7]).

We next consider the solutions with even symmetry (i.e., $I_{-m} = I_m$). Now $\mathbf{H}_0 \mathbf{I}^{(0)} = 0$ so long as $\sum_m I_m^{(0)} = 0$. This current distribution is such that its magnitude is zero at the gap. $\mathbf{I}^{(0)}$ is in general not a solution of the circuit equation because the internal emf inside the ring is not zero: $E_{\text{int}}(\phi) = \sum_m E_m \exp(im\phi) \neq 0$, where $E_m = X_m I_m^{(0)}$. The only way the circuit equation can be satisfied is if $E_{\text{int}}(\phi) \propto \delta(\phi)$; the internal emf is zero inside the ring except at the gap where it is counterbalanced by the gap resistance. For this to be true, it is necessary that all $E_m$'s be the same. We thus write $E_m = X_m I_m^{(0)} = E'$ and $I_m^{(0)} = E'/X_m$. Since $\sum_m I_m^{(0)} = 0$, we arrive at the eigenvalue equation $\sum_m 1/X_m = 0$. Substituting in the expression for $X_m$, we get the following equation for the entire spectrum of the even modes:

$$1 + \sum_{m=1}^{\infty} \frac{2(L_0 \omega^2 - ir_c\omega)}{L_m \omega^2 - m^2/C_m - ir_c\omega} = 0. \qquad (4)$$

For the low lying modes, contributions from terms with large $m$ are not important because of the $m^2$ coefficient multiplying $1/C_m$. The eigenmode only involves a few Fourier components, which agrees well with the eigenvector calculated numerically (See Fig. 3 of [7]).

We now describe a formal approach to solve the circuit equation, which can be easily extended to more complicated situations. This approach explicitly displays the eigenvectors, which can be used to calculate the responses of the structures to external fields. Because $r$ is very big, we solve the matrix problem (1) by a standard perturbation method. For the unperturbed matrix problem, $\mathbf{H}_0 \mathbf{I}^{(0)} = \mathbf{0}$, any vector $\mathbf{I}^{(0)}$ satisfying

$$\sum_m I_m^{(0)} = 0 \qquad (6)$$

is a solution. Now consider the full matrix problem Eq. (1). Assume the solution to Eq. (1) can be written as $\mathbf{I} = \mathbf{I}^{(0)} + \mathbf{I}^{(1)}$, where $\mathbf{I}^{(1)}$ is of the order of $r^{-1}\mathbf{I}^{(0)}$. Put $\mathbf{I}$ into Eq.

(1), since $\mathbf{H_0 I^{(0)}} = \mathbf{0}$, we get $\mathbf{HI} = (r\mathbf{M}+\mathbf{X})(\mathbf{I^{(0)}}+\mathbf{I^{(1)}}) = r\mathbf{MI^{(1)}} + \mathbf{XI^{(0)}} + o(r^{-1}) = 0$.
Substituting in the elements of $\mathbf{X}$ and $\mathbf{M}$, we get $I_m^{(0)} = E'/X_m$ where $E' = -r\sum_{m'} I_{m'}^{(1)}$ is a constant independent of $m$. Employing the constrain (6), we arrive at the equation $\sum_m 1/X_m = 0$, which again leads to Eq. (4). Choosing an appropriate normalization constant, in the limit $1/r \to 0$, we find the eigenvector at resonance to be

$$\mathbf{I} = \mathbf{I^{(0)}} = [..., X_0/X_2, X_0/X_1, 1, X_0/X_1, X_0/X_2, ...]^T. \tag{7}$$

We emphasize that Eqs. (4) and (7) are the *exact* solutions of the matrix problem (1), since the perturbation theory becomes *exact* in the limit of $1/r \to 0$.

### III. Double Ring

Realizing the difficulties of miniaturizing the resonance structure of a single-ring SRR, we now consider the double-ring system, in which one ring has a gap at $\phi = 0$ and the other one, $\phi = \pi$. In what follows, we develop analytical formulas suitable for the following two cases that were widely studied in literature: (1) the two rings are of slightly different size and located on the same plane [3-6,10]; (2) the two rings are of the same size and located on different planes separated by a distance $d$ with centers both on the $z$ axis [5]. For both cases, the circuit equation to determine the resonance frequencies is given by $\mathbf{HI} = (\mathbf{H_0} + \mathbf{Y})\mathbf{I} = 0$. The unperturbed and the perturbation matrices are

$$\mathbf{H_0} = \begin{pmatrix} r\mathbf{M^a} & 0 \\ 0 & r\mathbf{M^b} \end{pmatrix}, \quad \mathbf{Y} = \begin{pmatrix} \mathbf{X} & \mathbf{X'} \\ \mathbf{X'} & \mathbf{X} \end{pmatrix} \tag{8}$$

where the matrix $\mathbf{M^a}$ (with elements $\mathbf{M^a}_{m,m'} = 1$) is for ring 1 and the matrix $\mathbf{M^b}$ (with elements $\mathbf{M^b}_{m,m'} = (-1)^{m-m'}$) is for ring 2. $\mathbf{X}, \mathbf{X'}$ are diagonal matrices with $X_m$ same as before and $X'_m = i[L'_m \omega - m^2/(C'_m \omega)]$, in which $L'_m$ and $C'_m$ are the mutual inductances and capacitances between two rings. Here, we have neglected the size differences of the two rings in writing the self interaction terms. Again, there are even and odd modes under the transformation $m$ to $-m$. Here we only focus on the even modes. Following the perturbation theory, we first consider the unperturbed matrix problem $\mathbf{H}_0\mathbf{w} = \mathbf{0}$. The solutions can be written as $\mathbf{w} = \begin{bmatrix} \mathbf{w}^a \\ \mathbf{w}^b \end{bmatrix}$ which satisfy

$$\sum_m \mathbf{w}^a_m = 0, \quad \sum_m (-1)^m \mathbf{w}^b_m = 0. \tag{9}$$

Now assume the full solution as $\mathbf{I} = \mathbf{w} + \boldsymbol{\varepsilon}$ where $\boldsymbol{\varepsilon}$ is of the order of $1/r$, we get $\mathbf{HI} = \mathbf{Yw} + \mathbf{H}_0\boldsymbol{\varepsilon} + o(r^{-1}) = 0$. Putting Eqs. (8) into the above matrix and doing the algebra, we find

$$\begin{cases} \mathbf{w}^a_m = \left[ X_m C^a - (-1)^m X'_m C^b \right] / \left[ X_m^2 - X_m^{'2} \right] \\ \mathbf{w}^b_m = \left[ (-1)^m X_m C^b - X'_m C^a \right] / \left[ X_m^2 - X_m^{'2} \right] \end{cases} \tag{10}$$

where $C^a = -r \sum_{m'} \boldsymbol{\varepsilon}^a_{m'}$, $C^b = -r \sum_{m'} (-1)^{m'} \boldsymbol{\varepsilon}^b_{m'}$ are two constants independent of index $m$. Applying the constrains (9) to (10), we finally arrive at the equation $\sum_m \frac{1}{X_m \mp (-1)^m X'_m} = 0$ to determine the resonance frequencies of the double-ring system. We note that this equation recovers the single-ring results (i.e., Eq. (4)) in the absence of mutual-interaction terms (i.e., $X'_m = 0$). In explicit form, this equation becomes

$$\sum_{m=1}^{\infty} \frac{2\omega^2(L_0 \pm L'_0) - 2ir_c\omega}{i\omega r_c - L_m\omega^2 + m^2/C_m \pm (-1)^m \left[ -L'_m\omega^2 + m^2/C'_m \right]} = 1 \tag{11}$$

Again, we emphasize that Eqs. (11) is the *exact* solution of the double-ring problem (recalling $r \to \infty$).

Under the three mode approximation [retaining only the $m = 1$ term in Eq. (11)], we obtain, for $r_c = 0$, the resonance condition for the lowest (magnetic) mode (plus sign),

$$\omega_0^2 = (1/C_1 - 1/C_1')/(2L_0 + 2L_0' - L_1' + L_1). \tag{12}$$

The frequency squared is reduced because it is proportional to the *difference* between the inverse self capacitance $1/C_1$ and the inverse mutual capacitance $1/C_1'$. Physically this comes about because the charge distributions on the two rings are proportional to $\sin\phi$ and $-\sin\phi$ and nearly cancel each other. For rings made with wires of circular cross sections, such a difference is proportional to $\ln(d/a)$ [10], and the resonance frequency is reduced when two rings approach each other ($d$ decreases). Unfortunately, since the wires are of circular cross sections, the minimum value for $d$ is $2a$, so that the resonance frequency never approaches zero but possesses a natural lower bound [10]. This problem can be remedied if the rings are made with flat wires, and are placed on different planes [see inset to Fig. 1(a)]. We describe this next.

**IV. Non coplanar rings**

For rings made with flat wires not on the same plane, the separation $d$ between rings need only be larger than the thickness $t$ of each ring. In general $t$ can be much smaller than width $2a$ of the flat wires, so that the difference of capacitances, and in turn, the resonance frequency, can be much smaller.

Our conclusion remains valid when higher modes are included. To illustrate, when the $m = \pm 2$ modes are included, the resonance frequency is given by, for $r_c = 0$,

$$\omega_0'^2 = (1-\delta)(1/C_1 - 1/C_1') / \left[ 2(L_0 + L_0') + (1-\delta)(L_1 - L_1') \right]$$ where

$\delta = 2\omega_0^2 (L_0 + L_0')/(1/C_2 + 1/C_2')$. Thus as $1/C_1 - 1/C_1'$ is made small, the resonance frequency becomes small as well.

We now demonstrate the above argument by numerically evaluating the resonance frequencies of double stacking rings made with flat wires. As shown in the inset to Fig. 1(a), the planes of the rings are defined by an angle $\beta = \tan^{-1}(d/2R)$. Extending our previous theory for wires with circular cross sections [7,10] to the present case with rectangular cross sections [11], we are able to calculate all the circuit parameters $L_m, L_m', C_m, C_m'$ in terms of the geometrical parameters $d$, $R$, $a$, and $t$. For example, the self- and mutual- capacitance parameters are found to be

$$1/C_m = m^2 /(8\varepsilon_0 a^2) \sum_{l=|m|}^{\infty} (l-m)!/(l+m)! |P_l^m(0)|^2 \int_{R-a}^{R+a} d\rho/\rho \int_{R-a}^{R+a} d\rho' \rho_<^l / \rho_>^{l+1} \quad,$$

$$1/C_m' = m^2 /(8\varepsilon_0 a^2) \sum_{l=|m|}^{\infty} (l-m)!/(l+m)! P_l^m(-\sin\beta) P_l^m(\sin\beta) \int_{R-a}^{R+a} d\rho/\rho \int_{R-a}^{R+a} d\rho' r_<^l / r_>^{l+1}$$

, where $r = \sqrt{\rho^2 + (d/2)^2}, r' = \sqrt{\rho'^2 + (d/2)^2}$, $r_>(r_<)$ and $\rho_>(\rho_<)$ take the larger (smaller) values of $r, r'$ and $\rho, \rho'$, respectively. Similar expressions are found for $L_m, L_m'$. It is important to note that parameter $t$ does not enter the expressions of the circuit parameters in the thin-wire limit ($t \to 0$) [11].

We have numerically evaluated these circuit parameters, and found that in the limit of $a/R \to 0$, both $L_m$ and $1/C_m$ still exhibit the $\ln(R/a)$ dependences, similar to the circular cross-section case [7,10]. Thus the analytical results obtained for circular cross-section case, e.g., Eqs. (3,5), remain valid for the single-ring SRR made with flat wires. We next consider the double-ring case.

When both $a$ and $R$ are fixed, the normalized inverse-capacitance differences, $1 - C_m / C_m^{'}$, approach zero as $d \to 0$, as shown in Fig. 1(a) calculated for $a/R = 0.025$. As $d \to 0$, $\sin \beta \to 0, r \to \rho, r' \to \rho'$; thus $1/C_m^{'} \to 1/C_m$. As a result, the lowest resonance frequency ($\omega_0$) of the double-ring SRR is significantly reduced as $d \to 0$, as shown in Fig. 1 (b) by the solid line, obtained by numerically solving Eq. (11) setting $a/R = 0.025$ with the circuit parameters described above and similar expressions for $L_m$ and $L_m^{'}$ [11].

We performed finite-difference-time-domain (FDTD) simulations [12] on a series of realistic double-ring SRR's made by flat wires with different separation $d$. The FDTD calculated $\omega_0$ are shown in Fig. 1(b) as solid stars, which agree quite well with the analytic solutions. The small discrepancies between the FDTD and the analytic results can be attributed to the approximations adopted in calculating the circuit parameters [11].

However, FDTD simulations are difficult to perform for the very small $d$ cases, since in such cases the basic mesh discretizing the structure becomes too fine. Fortunately, analytic formulas are available when $d \to 0$. Expanding $r_<, r_>$ and $P_l^m (\sin \beta)$ as a power series in $d/R$ and keeping the lowest order terms, we find that $1 - C_1 / C_1^{'} \approx F \cdot (d/R)^2$, where $F$ is a dimensionless coefficient depending only weakly on $a/R$. Considering the 3-mode expression as shown in Eq. (12), we further find that $\omega_0 / \omega_u \approx \tilde{F} \cdot (d/R)$ with $\tilde{F}$ being another dimensionless parameter. Shown in Fig. 2 are $1 - C_1 / C_1^{'}$ and $\omega_0 / \omega_u$ as functions of $d/R$, calculated with the full theory setting $a/R = 0.025$. These numerical results accurately confirmed the above two formulas, and suggested that $\tilde{F} = 66$ and $F = 40240$ for the present structure.

This analytical formula enables us to estimate the lowest possible value of $\omega_0 / \omega_u$ in practical situations. Experimentally, it is possible to make bilayer systems separated by a distance of $d$ of the order of 10 nm. For rings of radius $R = 10$ mm and $a = 0.25$ mm, we find from the formula that $\omega_0 \approx 7 \times 10^{-5} \omega_u$, indicating that the longest resonance wavelength can be made of the order of $10^5$ the radius of the ring if $r_c$ can be ignored. This huge ratio is previously inaccessible. Experimentally both superconducting [13] and ordinary SRR's have been studied. For nonsuperconducting rings, a detailed examination of Eq. (11) under the three-mode-approximation shows that our estimate is valid (i.e., $\omega_0$ taking a non-zero real part) only when $r_c < (4/3)\sqrt{L_0/C_1}\left(1 - C_1/C_1'\right)^{0.5}$. This implies that the total resistance of the ring, $r_c \cdot 2\pi R$, has to be approximately less than $Z_0 g \left(1 - C_1/C_1'\right)^{0.5}$, where $Z_0 = (\mu_0/\varepsilon_0)^{1/2} = 377\Omega$ is the vacuum impedance, and $g = 2\sqrt{L_0/C_1}/3\pi R Z_0$ is a dimensionless constant, depending only weakly on $a/R$ (for $a/R = 0.025$, we get $g = 6.26$).

## V. Conclusions

In short, through consistently analytical, numerical and brute-force simulation studies, we have demonstrated how to create resonance structures whose sizes are much smaller than the resonance wavelength. The calculation described in this paper can be generalized to other structures, say, a four-ring system, which can further lower the resonance frequency, and can be extended to calculate analytically the responses of resonant structures to external field. STC is partly supported by a grant from the DOE. LZ thanks supports from the National Basic Research Program of China, the NSFC, Fok Ying Tung Education Foundation, and PCSIRT.

**Figures**

Fig. 1 (a) Differences between inverses of self- and mutual capacitances as the functions of ring-ring distance, for a double-ring SRR.  Inset schematically shows the geometry of a double-ring SRR, where the two shaded areas represent the cross sections of two flat wires. (b) Lowest resonance frequency of the double-ring SRR as a function of the ring-ring distance, calculated by the present analytical theory (line) and the FDTD simulations on realistic structures.

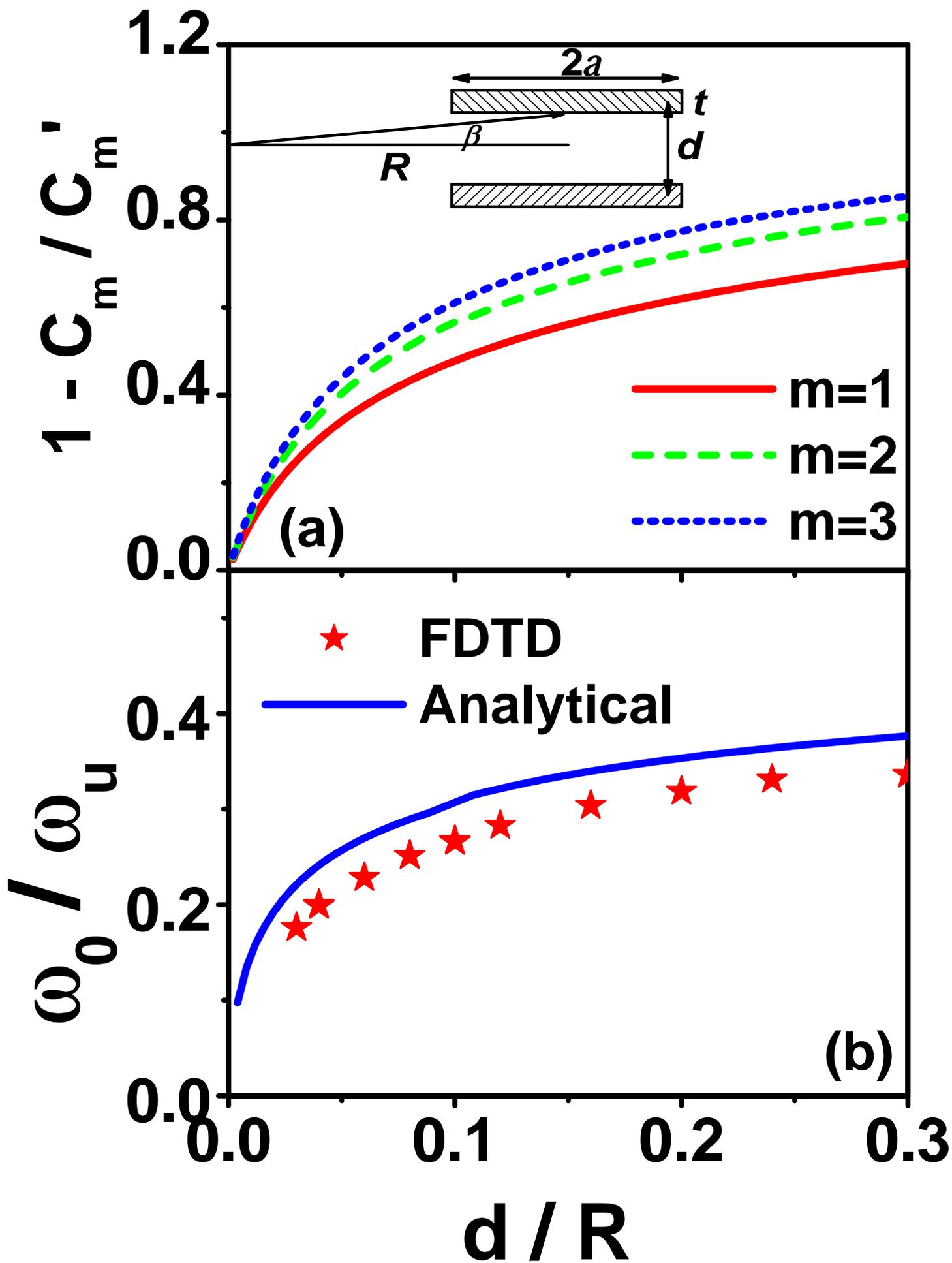

Fig. 2 $\omega_0/\omega_u$ and $1-C_1/C_1'$ as the functions of $d/R$ in the limit of very small $d/R$, calculated based on the analytical theory (symbols) for $a/R=0.025$, and the fitting formulas shown in the figure.

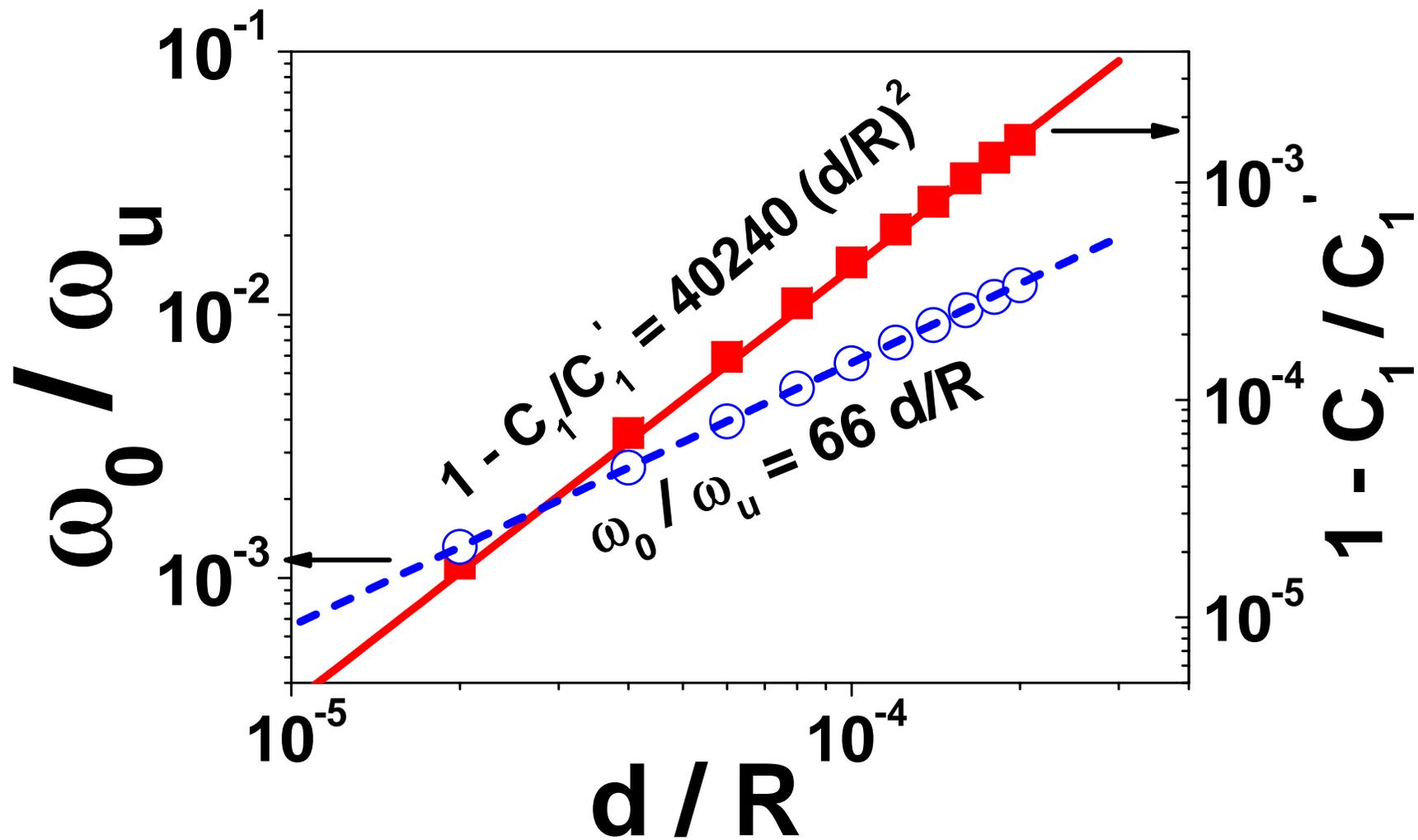